\documentclass[twocolumn,prX,nofootinbib]{revtex4}
\usepackage{amssymb}
\usepackage{enumerate}
\usepackage{footnote}
\usepackage{graphicx}

\begin{document}
\title{XAX: a multi-ton, multi-target detection system\\ for dark matter, double beta decay and pp solar neutrinos}

\author{K. Arisaka}
\author{H. Wang}
\author{P. F. Smith}
\author{D. Cline}
\author{A. Teymourian}
\author{E. Brown}
\author{W. Ooi}
\author{D. Aharoni}
\author{C. W. Lam}
\author{K. Lung}
\author{S. Davies}
\author{M. Price}
\address{Department of Physics and Astronomy, University of California, Los Angles, USA}

\begin{abstract}
A multi-target detection system XAX, comprising concentric 10 ton targets of $^{136}Xe$ and $^{129/131}Xe$, together with a geometrically similar or larger target of liquid Ar, is described.  Each is configured as a two-phase scintillation/ionization TPC detector, enhanced by a full $4\pi$  array of ultra-low radioactivity Quartz Photon Intensifying Detectors (QUPIDs) replacing the conventional photomultipliers for detection of scintillation light.  It is shown that background levels in XAX can be reduced to the level required for dark matter particle (WIMP) mass measurement at a $10^{-10}$ pb WIMP-nucleon cross section, with single-event sensitivity below $10^{-11}$ pb.  The use of multiple target elements allows for confirmation of the $A^2$ dependence of a coherent cross section, and the different Xe isotopes provide information on the spin-dependence of the dark matter interaction. 
The event rates observed by Xe and Ar would modulate annually with opposite phases from each other for WIMP mass $>\sim$100 GeV$/c^2$.  The large target mass of $^{136}Xe$ and high degree of background reduction allow neutrinoless double beta decay to be observed with lifetimes of $10^{27}$-$10^{28}$ years, corresponding to the Majorana neutrino mass range 0.01-0.1 eV, the most likely range from observed neutrino mass differences. The use of a $^{136}Xe$-depleted $^{129/131}Xe$ target will also allow measurement of the pp solar neutrino spectrum to a precision of 1-2\%.   
\end{abstract}

\maketitle

\section{Overall description of detection system}
This paper investigates the properties of the proposed multi-ton detector XAX\footnote{Acronym for ``$^{129/131}Xenon-Argon-^{136}Xenon$'', pronounced ``ZAX''} comprised of three noble liquid targets. XAX is designed to detect nuclear recoils from dark matter, events from neutrinoless double beta decay, and pp solar neutrino scattering events. The target materials are:

\vspace*{3mm}
\noindent Target 1: Liquid Xenon depleted in $^{136}Xe$ (enriched in $^{129/131}Xe$) for:

\newenvironment{my_enumerate}{
\begin{enumerate}[(i)]
  \setlength{\itemsep}{1pt}
  \setlength{\parskip}{0pt}
  \setlength{\parsep}{0pt}}{\end{enumerate}
}
\newenvironment{my_enumerate1}{
\begin{enumerate}[(a)]
  \setlength{\itemsep}{1pt}
  \setlength{\parskip}{0pt}
  \setlength{\parsep}{0pt}}{\end{enumerate}
}

\begin{my_enumerate}
\item dark matter interactions with nuclei of non-zero spin.
\item pp solar neutrino interactions (via $\nu_{e}$ scattering).
\end{my_enumerate}

\noindent Target 2: Liquid Xenon enriched in $^{136}Xe$ for:

\begin{my_enumerate}
\item dark matter interactions with zero-spin nuclei.
\item neutrinoless double beta decay.
\end{my_enumerate}

\noindent Target 3: Liquid Argon depleted in $^{39}Ar$ (or optionally liquid Neon) for:
\begin{my_enumerate}
\item comparison of dark matter interactions in Xenon with lower-A dark matter target.
\item additional pp solar neutrino target (in the case of Ne).
\end{my_enumerate}

\vspace*{3mm}
A possible configuration is sketched in Fig 1, consisting of two cylindrical, active target volumes of 2 m diameter and 2 m height: (a) 10 tons liquid $^{129/131}Xe$, surrounding 10 tons of liquid Xe enriched in $^{136}Xe$, and (b) 10 tons (or larger) liquid Ar in a separate vessel. Two cryostats are located within a single shielding enclosure which provides purified water shielding thicker than 4 meters in all directions.  Events in the water are separately monitored by photomultipliers to provide an active Cerenkov veto against the residual underground muon flux.

\begin{figure}
\includegraphics[height=55mm]{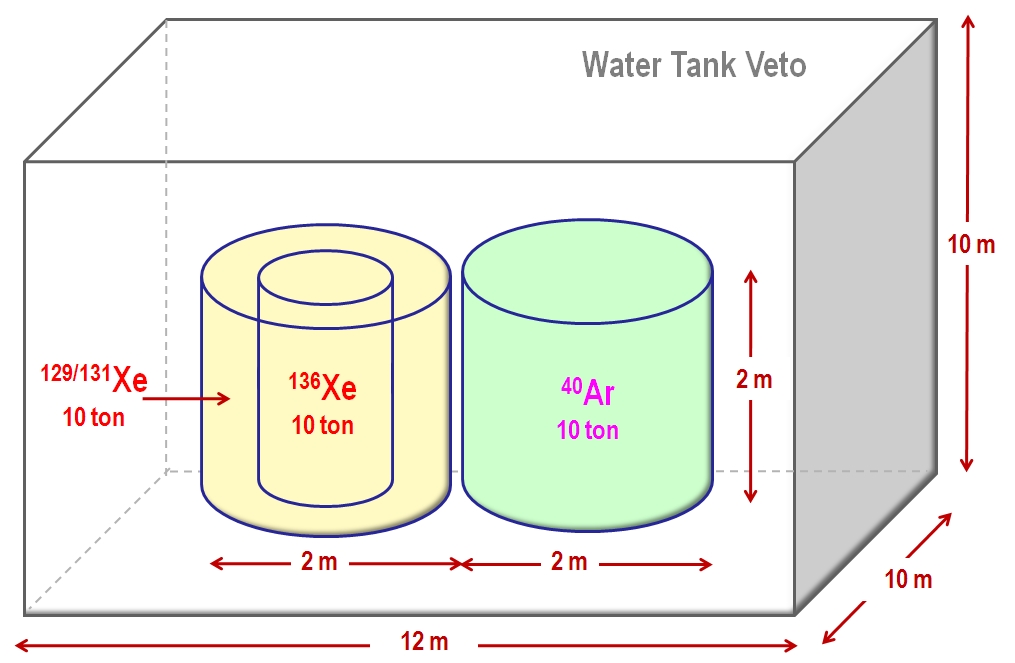}
\caption{\label{fig:epsart} Conceptual view of XAX detector layout.}
\end{figure}

\vspace{3mm}
The expected energy spectra for the three scientific objectives (dark matter, double beta decay and pp solar neutrinos) are shown in Fig 2. For simplicity, natural Xe is assumed here as a target material. In addition, 100 GeV/$c^{2}$, 1 and 10 TeV/$c^{2}$ WIMP masses with $10^{-8}$ pb $(=10^{-44} cm^{2})$ of nuclear cross section and two-neutrino/neutrinoless double beta decays with life times of $10^{22}$/$10^{27}$ years are assumed and plotted together with $pp/^{7}Be/^{8}B$ solar neutrino spectra.

\begin{figure*}
\includegraphics[height=90mm]{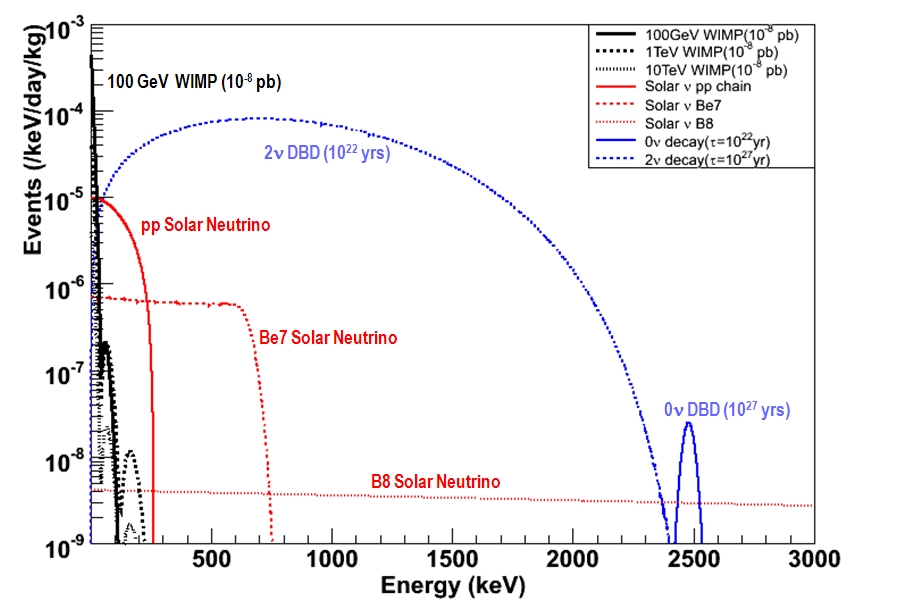}
\caption{\label{fig:epsart} Expected energy spectrums of WIMPs, solar neutrinos, and double beta decays in case of natural Xe as a target material. 100 GeV/$c^{2}$, 1 and 10 TeV/$c^{2}$ WIMP masses of WIMP with $10^{-8}$pb nuclear cross section; $2\nu \beta \beta$/$0 \nu \beta \beta$ with life times of $10^{22}$/$10^{27}$ years are assumed. The observed energy is smeared out assuming a resolution of  $\sigma=\frac{1.5 \%}{ \sqrt{ E \left( MeV \right)}}$.}
\end{figure*}

Detection of dark matter events is based on the well-established two-phase TPC (Time Projection Chamber), giving both primary (S1) and secondary (S2) scintillation signals for each event, from which nuclear recoils can be discriminated from gamma background with less than 1\% overlap \cite{Aln,Ang,Ben}.  This method currently reaches a sensitivity  of $< 10^{-7}$ pb as shown in Fig 7. The purpose of this paper is to demonstrate that a factor of 1000 improvement on the existing WIMP sensitivity can be achieved in XAX by the use of larger target masses with improved self-shielding and a new ultra-low background photon detection device QUPID surrounding each target volume in a $4\pi$  array for position sensitivity.  The same scintillation photon detection system will also be capable of recording events at 2479 keV from neutrinoless double beta decay in $^{136}Xe$, as well as the spectrum of 0-250 keV electron recoils from pp solar neutrino scattering in Xe.

The basic detection principles of the double phase TPC have been fully demonstrated on a smaller scale in previous/ongoing projects with both Xe \cite{XeTPC} and Ar \cite{ArTPC}. In this paper we describe, for each type of signal, how backgrounds can be reduced in much larger target masses to the new low levels needed. The most important problem, common to the noble liquid detectors, is the reduction of gamma and neutron backgrounds from the photodetector array.  To achieve this, we are developing a new device - a Quartz Photon Intensifying Detector, or QUPID, which we anticipate will have a U/Th impurity level (and hence gamma and neutron emission) a factor of $\sim100$ lower than that of the current best low-radioactive photomultipliers. 

In addition to the ionization/scintillation signal discrimination provided by a double phase TPC, the Argon portion of the detector will also employ pulse shape discrimination in order to improve the efficiency of determining nuclear recoils versus $^{39}Ar$ beta decay events. The difference between the two is based upon the singlet and triplet dimer states of liquid Argon, each of which has a unique lifetime. Thus, the timing information of the pulse provides a method for identifying nuclear and electronic events. Further information can be found in \cite{PSD}.

\section{Conceptual detector design}
A conceptual detector design of a single vessel is shown in Fig 3 with a $4\pi$ photodetector array. A large transparent vessel for the active target material (as well as a cylindrical container for $^{136}Xe$) is made from low-radioactive Acrylic, similar to the heavy-water Acrylic vessel developed by SNO \cite{SNO}. The internal wall is painted with a Wave Length Shifter (WLS, such as TPB) to convert the UV scintillation photons to visible light, similar to the technique developed by WARP \cite{WARP1} and DEAP/CLEAN \cite{CLEAN1}.  Furthermore, to provide uniform vertical electric fields for drifting ionized electrons upwards, the internal cylindrical walls are either painted by a transparent resistive layer, such as ATO (Antimony-Tin-Oxide) as in Fig 4, or coated by field shaping electrodes created by fine aluminium patterns. The inner surface of the bottom plate of the Acrylic vessels is painted with a combination of WLS and a thin transparent conducting layer, such as ITO (Indium-Tin-Oxide) or Platinum coating.  

\begin{figure}
\includegraphics[height=80mm]{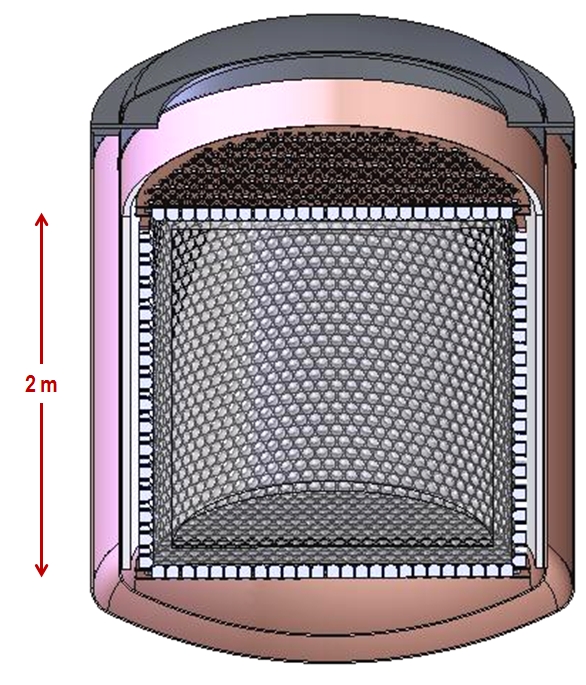}
\caption{\label{fig:epsart} A cross-sectional view of one of the XAX detectors, showing a double-layer copper cryostat, which contains noble liquid (Xe, Ar or optionally Ne), surrounded by a $\sim$10 cm-thick acrylic vessel (with TPB coating) and closely packed 3950 units of 3 inch diameter QUPIDs.}
\end{figure}

\begin{figure}
\includegraphics[height=80mm]{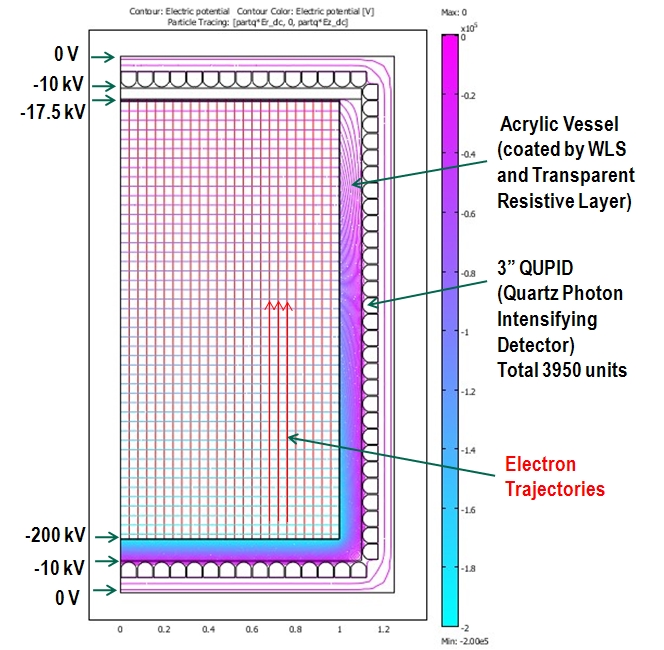}
\caption{\label{fig:epsart} Simulation of electric field lines created by a current through the transparent resistive layer to drift electrons in liquid Xe/Ar vertically to the Xe/Ar gas. Application of a constant current along the side wall produces an extremely uniform potential gradient, creating perfectly horizontal equipotential planes.}
\end{figure}

\begin{figure*}
\includegraphics[height=68mm]{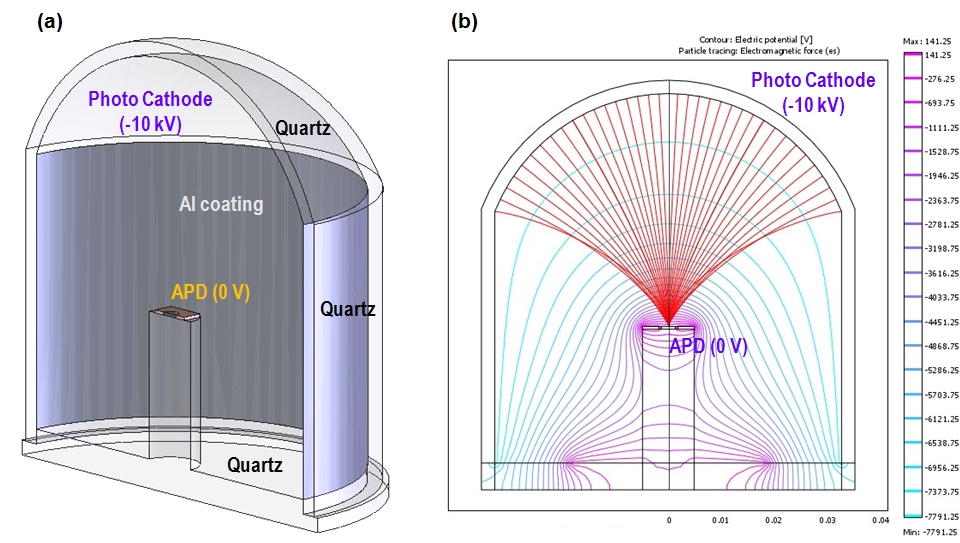}
\caption{\label{fig:epsart} (a) 3D cut-view of a 3 inch diameter QUPID, (b) simulated electric field lines and electron trajectories. -10 kV is supplied on the photocathode and photoelectrons are focused onto the center of the  APD which is set to a ground level.}
\end{figure*}

The outside of the Acrylic vessel is completely surrounded by a closely-spaced $4\pi$ array of low background photo detectors (QUPIDs). This $4\pi$ arrangement not only provides high detection efficiency of primary scintillation light (S1), but also measures its position with an accuracy of 10-20\% of the detector diameter, giving a 3-D position sensitivity better than 1\% of the detector volume. From this point of view, XAX takes full advantage of conventional single-phase detectors with $4\pi$ geometry (such as XMASS \cite{XMASS} and DEAP/CLEAN \cite{CLEAN}) in terms of S1 signal detection. In the case of XAX, however, the benefits of double-phase detectors are also realized. S2 signals, obtained from the TPC operation, can provide 3-D position reconstruction accurately, as is the case in conventional double-phase detectors (such as ZEPLIN-II \cite{Aln}, XENON10 \cite{Ang}, and WARP \cite{WARP}). What makes XAX unique is the possibility to impose 4-D (position and time) coincidence between the S1 and S2 signals for additional reduction of background.

The whole structure is housed in a double-layer vacuum cryostat made of a low radioactive metal such as OFHC (Oxygen-Free High Conductivity) Copper as shown in Fig 3.  It is then immersed in a large water tank as shown in Fig 1.

\section{Low background light detection - QUPID}
Existing dark matter detectors using liquid Xe or liquid Ar can remove most gamma and neutron backgrounds through shielding and veto techniques, but a residual background remains, in particular from photomultipliers, which are located close to the target. This background arises principally from U and Th impurities in the photomultiplier components, giving both a gamma background directly from decay of the U/Th chains, and a neutron background from alpha-n production.  Over the past 10 years, improvements in radiopurity have reduced these by two orders of magnitude to the level of $2\times10^{3}$ gammas/day and $2\times10^{-3}$ neutrons/day per 5 cm PMT such as Hamamatsu R8778 \cite{Ben}.  

To further reduce background by another factor of 100 (paving the way for the ultra-low background $4\pi$  arrays of photon detectors required for XAX), a design for an all-quartz Hybrid Avalanche Photodiode (HAPD), called the QUPID (Quartz Photon Intensifying Detector) has been proposed by K. Arisaka and H. Wang. Advantages of the HAPD over a conventional photomultiplier are summarized in \cite{Ari1} and technical details of the QUPID can be found elsewhere \cite{Ari2}. Briefly, the key features of the QUPID are shown in Fig 5. The whole structure is made by ultra-pure synthetic fused silica except for the very small avalanche photodiode (APD) at the center and its readout leads. Photoelectrons are emitted by the hemispherical photocathode and accelerated by 10 kV to the APD, producing $\sim2,000$ secondary electrons per photoelectron by direct bombardment,  which are then avalanched within the APD by a further factor $\sim30$, giving a total gain of $\sim60,000$.  The electron trajectory simulation in Fig 5(b) shows that photoelectrons are well-focused towards the central region of the APD. 

The unit, currently being developed by UCLA and Hamamatsu Photonics (Hamamatsu, Japan), has an outside diameter of 70 mm, with a photocathode diameter of 65 mm.  This all-quartz design should achieve no more than 10-100 emitted gammas and 0.01 - 0.1 emitted neutrons/year from each individual QUPID.  Note that no resistor/capacitor chain is required, so  additional radioactivity from these components is eliminated.  The quantum efficiency is expected to be 30-35\% (at 170-450 nm wavelength), with an anticipated yield of approximately 4 photoelectrons/keV for a $4\pi$  array surrounding a 2 m diameter detector volume.  To ensure good photocathode dynamic range with superior linearity at the low temperature of liquid Ar and Ne, a special photocathode material is being developed by Hamamatsu Photonics and applied to the QUPID. Combined with this new photocathode,  the intrinsically superior dynamic range of the HAPD permits an extremely wide range of signal detection from $\sim5$ keV dark matter signals to 2.5 MeV double beta decay signals.

\section{Galactic dark matter (WIMP) detection}
The experimental requirements for galactic particle dark matter detection and identification follow from the expected differential energy spectrum of dark matter (WIMP) interactions with nuclei.   Direct search experiments are based on the assumption of a local mass density 0.3 GeV/$cm^{3}$ of hypothetical neutral particles of mass $M_{D}$ GeV and an average velocity $v_{0} = 220 km\cdot s^{-1}$.   If dark matter particles are incident on a nucleus of atomic number $A$, producing a recoil energy $E_{R}$ keV, the differential energy spectrum of the event rate $R$ (events/kg/day) is given by \cite{Lew}  

\begin{equation}\frac{dR}{dE_{R}}=\left(\frac{c_{1}R_{0}}{E_{0}r}\right)exp\left[\frac{-c_{2}E_{R}}{E_{0}r}\right]F^{2}\left(E_{R},A\right)\end{equation}

where $E_{0} = 0.5 M_{D}\frac{v_{0}}{c^{2}}\:keV$, $r =  \frac{4 M_{D}M_{T}}{\left(M_{D}+M_{T}\right)^{2}}$, $M_{T} =  0.932A$,  and $F^{2}$ is a nuclear form factor correction discussed in \cite{Lew}. The numerical constants $c_{1}$ and $c_{2}$ depend on the motion of the Earth-Sun system through the Galaxy; for a detector at rest with respect to an isotropic dark matter flux, $c_{1}$ = $c_{2}$ = 1, whereas motion through the Galaxy gives fitted values $c_{1} = 0.75$, $c_{2} = 0.56$, with a small modulation tabulated in \cite{Lew}. Fig 6 shows the energy spectra derived from this formula for several target materials being considered for WIMP search.  Here, a cross section of $10^{-8}$ pb is assumed for five target elements: liquid $^{132}Xe$, $^{40}Ar$, and $^{20}Ne$ (proposed in XAX), and solid $^{73}Ge$ and $^{28}Si$.

\begin{figure}
\includegraphics[height=66mm]{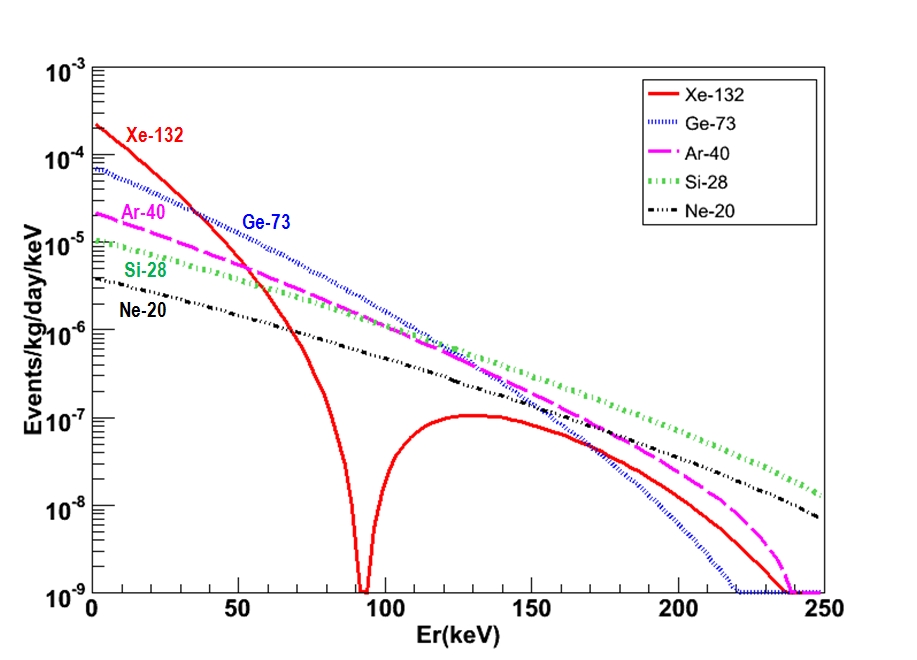}
\caption{\label{fig:epsart} Differential energy spectra of various target masses assuming a WIMP mass of 100 GeV. At lower recoil energies, Xenon is the preferred detection target because of its increased event rate.}
\end{figure}

\begin{figure*}
\includegraphics[height=100mm]{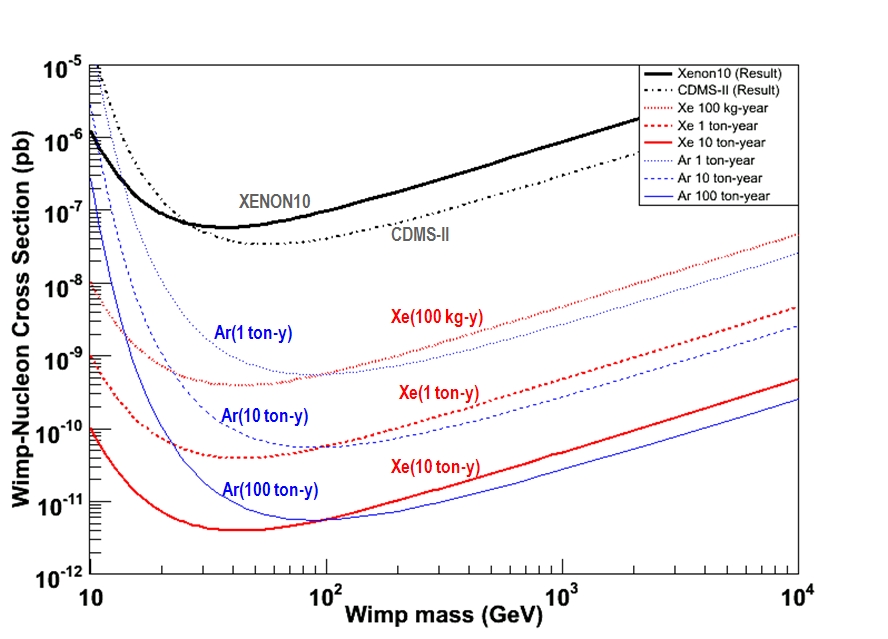}
\caption{\label{fig:epsart} 90\% Confidence Limit of WIMP-Nucleon cross section expected from Xe and Ar detectors with one year data taking. As a reference, the current best limits set by XENON10 and CDMS-II are plotted together.}
\end{figure*}

From Equation (1) the experimental measurement of, or limits on, the differential rate $\frac{dR}{dE_{R}}$  gives a value of, or limit on,  the quantity $R_{0}$ , which is defined as the total rate for a stationary Earth.  This is related to the total nuclear cross section  $\sigma_{A}$[pb]  by
\begin{equation}\frac{R_{0}}{r}=\frac{D\sigma_{A}}{{\mu_{A}}^{2}}\end{equation}

where $\mu_{A}=\frac{M_{D}M_{T}}{M_{D}+M_{T}}$  is the reduced mass of the colliding particles and $D$ is a numerical factor equal to 94.3 for an assumed dark matter density of 0.3 GeV/$cm^{3}$  \cite{Lew}.

Final limits are expressed in terms of the WIMP- nucleon cross section  $\sigma_{W-N}$[pb] given by
\begin{equation}\sigma_{W-N}=\left(\frac{\mu_{1}}{\mu_{A}}\right)^{2}\frac{\sigma_{A}}{A^{2}}\sim0.01\left(\frac{1}{A}\right)^{2}\left(\frac{R_{0}}{r}\right)\end{equation}

where $\mu_{1}$ ($\sim$~0.92 GeV) is the reduced mass for $A = 1$ and  $M_{D} > 50$ GeV. The coherence factor $A^{2}$ applies in the (expected) case of a predominantly spin-independent interaction, but would be absent in the case of an interaction only with a nucleon of unpaired spin.

To demonstrate the possibilities in a more quantitative way, MC simulations have been performed and yielded the results shown below.

\begin{figure*}
\includegraphics[height=73mm]{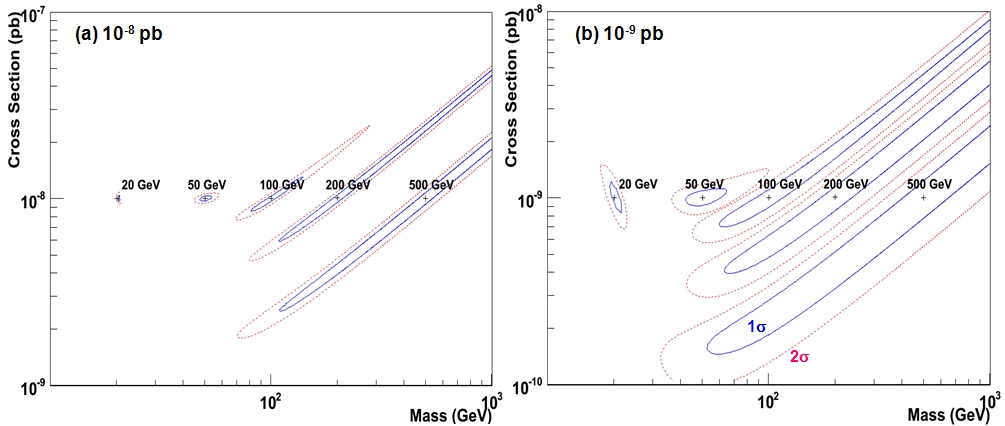}
\caption{\label{fig:epsart} $1\sigma$ and $2\sigma$ uncertainties in determining WIMP mass and cross-section by Xe target in XAX in 10 ton-years of data taking. WIMP masses 20, 50, 100, 200, 500 GeV, and cross sections of (a) $10^{-8}$ pb, (b) $10^{-9}$ pb  are assumed. Above 100 GeV/$c^{2}$ WIMP mass, there is degeneracy between mass and cross section, resulting in the $45^{\circ}$ slope in this figure.}
\end{figure*}

\subsection{Discovery potential}
With the improvements in background reduction possible with the XAX system, dark matter sensitivity down to $10^{-11}$ pb $(=10^{-47} cm^{2})$ is attainable (shown in Fig 7).  This should be compared with the current cross section limits (2008) at the level $10^{-7}$ - $10^{-8}$ pb \cite{Aln,Ang}, and theoretical expectations (from supersymmetry theory) extend to the level $10^{-10}$ pb and below \cite{SUSY}.  Thus, XAX could cover the most favored phase space predicted by supersymmetry models, if the neutralino is indeed the dark matter particle.

The order of magnitude difference in sensitivity between the Xe and Ar curves, shown in Fig 7, is principally the $A^{2}$ coherence factor in the WIMP mass range above 100 GeV/$c^{2}$. However, below 100 GeV/$c^{2}$, an even larger loss of sensitivity in Ar arises from its higher energy threshold ($\sim$30 keV) coming from large photoelectron statistics required by pulse discrimination. To compensate for such an effect, it would be beneficial to make the Ar detector an order of magnitude larger than the Xe detector in XAX, resulting in a total mass of 100 tons. Such enlargement may be feasible since Ar is several orders of magnitude less expensive than Xe.

\subsection{Confirmation of $A^{2}$ dependence of cross section}
Considerable effort will be made to eliminate background events which imitate recoils from dark matter: for example neutron scattering, radon decay nucleus recoils, or spurious data events.  However, unambiguous signal identification can be achieved only by candidate signals from targets of differing A, using the $A^2$ dependence in Equation (3).   By using liquid Ar as a second target material, the event rate (events/kg/day) will be reduced by an order of magnitude as shown in Fig 7.  For identification purposes, it is sufficient that the target masses of Xe and Ar are similar (eg 10 tons each), though a larger mass of Ar would provide an event rate large enough for an independent calculation of the WIMP mass. A further option is to replace the Ar with Ne for additional confirmation of the $A^{2}$ effect (and to provide an additional pp solar neutrino target, see \S7). One should note however that in the case of liquid Ne, double phase operation is not feasible because of the slow electron drift velocity.

\subsection{Determination of WIMP mass and nuclear cross section}
The cross section, predicted by the so-called focus point in minimal supersymmetry, is of order $10^{-8}$ pb \cite{SUSY}. Therefore, it is not unreasonable to assume this order of cross section. If so, 100-1000 events can be detected by 10 ton-years of XAX operation, in particular, by the Xe target. Taking this high statistics observation, from the shape of the energy spectrum, the WIMP mass can be deduced directly.  

Fig 8 shows $1\sigma$ and $2\sigma$ uncertainties in determining WIMP mass and cross-section with the Xe target in XAX in 10 ton-years of data taking, assuming (a) $10^{-8}$ pb and (b) $10^{-9}$ pb. Above 100 GeV/$c^{2}$ WIMP mass, there is a degeneracy between the mass and cross section, resulting in the $45^{\circ}$ slope in this figure.

\begin{figure*}
\includegraphics[height=65mm]{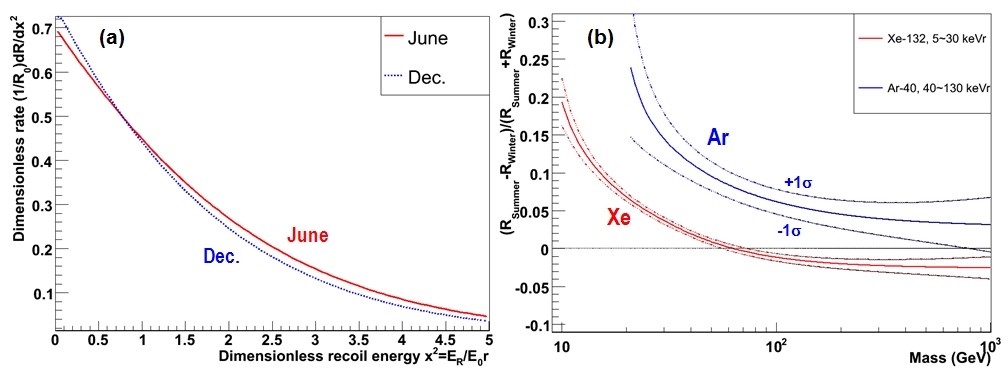}
\caption{\label{fig:epsart} (a) June and December dark matter energy spectra plotted in a dimensionless form. The curves cross at approximately a recoil energy $E_{R}\sim0.78 E_{0}r$  so that for data to the left of this point the annual modulation is reversed. (b) Normalized summer/winter difference as a function of particle mass, for both Ar and Xe targets, with  $\pm1\sigma$  errors for a 10 ton-year running period assuming a cross section of $10^{-8}$ pb.}
\end{figure*}

\subsection{Investigating spin-dependence}
Based on the hypothesis that the dark matter particle is the lightest particle of supersymmetry theory, its interaction may be a combination of spin-dependence and spin-independence, but the latter is always likely to dominate the cross section because of the multiplying factor $A^{2}$. Nevertheless there is interest in setting limits separately for spin-dependent and spin-independent interactions.  In the case of Xe this can be done by separating those isotopes with spin from those without spin.  For example, natural Xe is 48\% odd-A isotopes with spin 3/2 or 1/2 ($^{129}Xe$: 26.4\%, $^{131}Xe$: 21.2\%), and 52\% even-A isotopes with spin 0 ($^{132}Xe$: 26.9\%, $^{134}Xe$: 10.5\%, $^{136}Xe$: 8.9\%).  For a target enriched in the isotopes 132 and above, one could obtain a target consisting dominantly of spin zero nuclei.  An important special case is to enrich specifically $^{136}Xe$, which is a candidate for neutrinoless double beta decay, one of the three objectives of the XAX project.  This target isotope thus has the potential to provide not only a $0\nu\beta\beta$ signal, but also a dark matter signal for a spin-zero nucleus. This signal will be compared with the signal rate from a Xe target depleted in isotopes $\geq$132 in order to search for any rate difference related to the nuclear spin.  

\subsection{Observation of annual modulation}
The annual modulation effect (arising from the Earth's orbital motion superimposed on the Sun's galactic motion) can in principle prove the signal to be correlated with motion in the Galaxy. This requires stable operation over a period of years, together with full shielding of varying neutron and radon backgrounds. The difference in total rate between June and December would be no more than 4\%, but taking into account the variation in energy spectrum, a combined parameter would vary by up to 18\% \cite{Smith}.

Fig 9(a) shows the June and December dark matter energy spectra plotted in a dimensionless form. The curves cross at approximately a recoil energy $E_{R} \sim 0.78 E_{0}r$ so that for data to the left of this point the annual modulation is reversed. For a particle mass $\sim$100 GeV, the cross over point occurs at a recoil energy $\sim 8$ keV for an Ar target, and $\sim$ 32 keV for a Xe target. As a result, the majority of experimental data for Ar will be to the right of the crossover, and the majority of experimental data for Xe will be to the left of the crossover, as indicated by the ranges shown on Fig 9(b). The sign of the annual modulation will then be opposite for the majority of Xe data to that of the majority of Ar data. Thus extended operation of both Xe and Ar targets, and observation of both signs, would provide an interesting cross check on the annual modulation expectations, and an additional estimate of the dark matter particle mass.

Fig 9(b) shows the normalized summer/winter difference as a function of particle mass for both Ar and Xe targets, together with $\pm 1 \sigma$ errors for a 10 ton-year running period assuming a cross section of $10^{-8}$ pb. The data energy range used is that which is currently typical for the two target materials. The plots show the feasibility of estimating particle mass from the summer-winter event rates, and also the expected sign reversal above 100 GeV mass for the typical energy range of Ar data. It remains the case that with either target, one would see both signs of summer-winter difference in an experiment capable of covering the full energy range, as is clear from Fig 9(a).

To observe a statistically significant effect over 2 years, we would need observation of a total of 800 events, or 400/year, which requires a total of 20 tons liquid Xe at a $10^{-10}$ pb signal strength, or 2 tons at $10^{-9}$ pb.

\section{Background estimate}

To achieve a reasonable event rate at WIMP interaction cross sections down to $<1\times10^{-10}$ pb, the target mass has to be increased to the 10 ton scale proposed here, three orders of magnitude larger than the $\sim10$ kg scale of today's experiments. At the same time, the background rate/kg/day has to be reduced by a factor of 1000 to match the factor of 1000 lower signal rate. There are four major classes of backgrounds to be considered: 

\begin{my_enumerate1}
\item radiation from photon detectors (primarily from U/Th decay chains).
\item	pp solar neutrinos.
\item impurities in the noble liquid such as Kr and Rn, or internal radioactivity such as
$^{39}Ar$ (in case of Ar) or $^{136}Xe$ (in case of Xe, due to two-neutrino double beta decays).
\item radiation from outside the detector, such as low energy gammas/neutrons from shielding materials and rocks,
and	high energy gammas/neutrons induced by cosmic ray muon interaction in rocks.
\end{my_enumerate1}

In the following section, it is shown that the background level can be reduced by more than a factor of 1,000, to the level of $<$ 0.1 event per 10 ton-years, by combining (a) an ultra-low radiation photon detector, QUPID, (b) a 10 cm-thick active self-shielding cut by Liquid Xe/Ar, (c) a so-called S2/S1 cut for electromagnetic backgrounds (i.e. gamma rays and electrons), and (d) a multiple scattering cut in particular to reject neutrons.

\subsection{Radiation from photon detectors}

\subsubsection{Gamma backgrounds}
The pair of signals in a two-phase noble liquid TPC, referred to as S1 (direct scintillation in liquid) and S2 (proportional scintillation from the ionization avalanche in gas), currently produces a gamma rejection factor of greater than a hundred (at 50\% nuclear recoil efficiency) \cite{Ang}. 

As stated in \S3, the most difficult gamma background to remove is that from U/Th in the nearby photodetectors.  For conventional PMTs such as the Hamamatsu R8778, estimated target rates in the dark matter energy range are $\sim$10/kg/day for a 10 ton detector array. This reduces to $\sim10^{-3}$ /kg/day after a multiple scattering cut (since the weakly interacting WIMPs can scatter only once in this volume) and a 10 cm layer of self-shielding; the background reduces by an additional factor of 100 from an S2/S1 cut to $\sim10^{-5}$ /kg/day, just achieving the background level for detection of a $10^{-10}$ pb WIMP-nucleon cross section, but still masking a signal at $10^{-11}$ pb.  

The development of QUPIDs would gain a factor of 100 in reduction of gamma emission, and thus guarantee achievement of the backgrounds needed for measuring a signal at $10^{-10}$ - $10^{-11}$ pb WIMP-nucleon cross section. To further investigate the effectiveness of QUPIDs, extensive GEANT4 simulation has been performed, assuming conservative radioactivity of 1 mBq per QUPID and natural Xe. The obtained energy spectra from gamma ray interaction in the fiducial region of liquid Xe are plotted in Fig 10. Here the self-shielding cuts of 0 cm, 5 cm and 10 cm are applied after a multi-scattering cut and an S2/S1 cut (which is assumed have a rejection efficiency of 99\%). Once the 10 cm self-shielding cut is applied, the gamma ray background from QUPID is reduced down to $10^{-8}$ events/keV/kg/day, which is far below the level of pp solar neutrinos (the pp solar neutrino spectrum, which would become a dominant background at lower cross section levels, will be described later).

\begin{figure*}
\includegraphics[height=90mm]{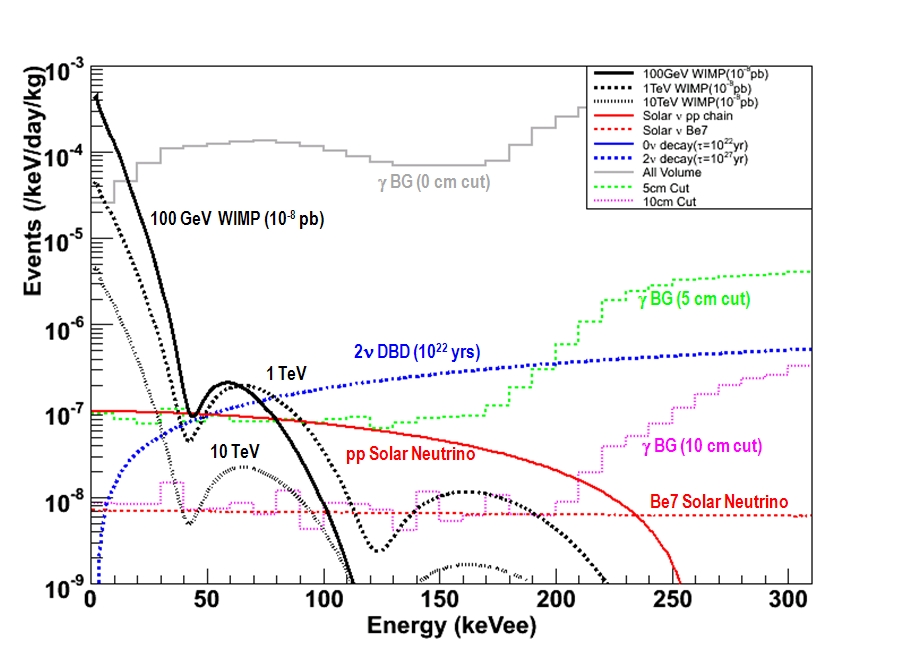}
\caption{\label{fig:epsart} Expected energy spectrum of WIMP interactions, solar neutrinos, double beta decays, and gamma ray backgrounds from QUPID as a function of self-shielding cuts. Natural Xe is assumed and both the S2/S1 cut and the multi-scattering cut have been applied in advance.  Note that the unit of energy in this figure is keVee (= electron equivalent energy).}
\end{figure*}

\begin{figure*}
\includegraphics[height=90mm]{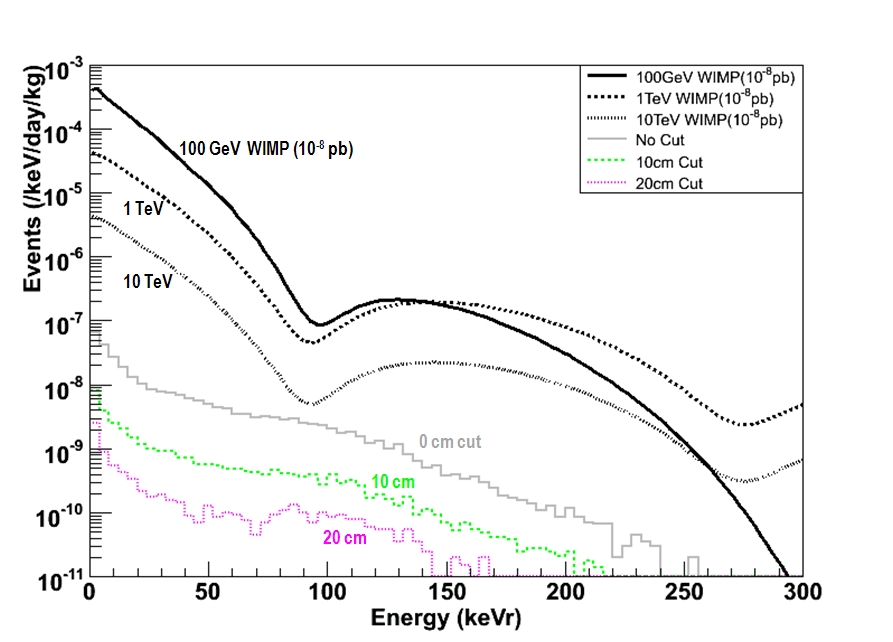}
\caption{\label{fig:epsart} Expected energy spectrum of WIMP, and neutron backgrounds from QUPID as a function of self-shielding cuts. Natural Xe is assumed and the multi-scattering cut has been applied in advance. Note that unit of energy in this figure is keVr (= nuclear recoil energy).}
\end{figure*}

\subsubsection{Neutron backgrounds}
To observe a positive signal at the $10^{-10}$ pb level, we require the neutron background to be reduced to at least an order of magnitude below the signal level (i.e. to $<10^{-5}$ events/kg/day).  

The major neutron background arises from U/Th in photodetectors and detector vessels through alpha-n production. Simulations based on the activity in R8778 PMTs and assuming 0.1ppb U/Th in nearby materials give neutron background rates in the region $10^{-4}$/kg/day \cite{Bun}, but this is reduced by a factor of 100 after a multiple scattering cut and a 10 cm outer layer of self-shielding in the target.  As discussed in \S3, a further factor of 100 decrease in photodetector activity would result from the use of QUPIDs in place of PMTs.  Thus the neutron background is reducible to well below the level required for the detection of events at $10^{-10}$ to $10^{-11}$ pb WIMP-nucleon cross section.

Again, to investigate the effectiveness of QUPIDs, extensive GEANT4 simulations for neutron interactions have been performed, assuming radioactivity of 1 mBq of U/Th per QUPID. The obtained energy spectra from neutron interaction in the fiducial region of Liquid Xe are plotted in Fig 11. Here the self-shielding cuts of 0 cm, 10 cm, 20 cm, and 30 cm are applied after a multi-scattering cut.  Once the 10 cm self-shielding cut is applied, the neutron backgound from QUPIDs becomes below $10^{-8}$ events/keV/kg/day, which is completely negligible.

\vspace{3mm}
Lastly, to illustrate the effectiveness of active self-shielding more graphically, interaction points of gamma ray and neutron backgrounds from QUPIDs are plotted in Fig 12 for a 2 m diameter and 2 m high target volume. All the plots are made by GEANT4 after imposing the energy window of the WIMP signal range, corresponding to 100-year-long data taking. 
 
Fig 12(a) shows the gamma backgrounds from U/Th decays after a multiple scattering cut and an S2/S1 cut in the case of Xe.  Figs 12(b) and 12(c) show the neutron background from U/Th decays after a multiple scattering cut in the cases of Xe and Ar respectively. Remarkably, once the self shielding cut of $\sim$10 cm thickness is imposed, there would be only a few background events even for 100-year-long data taking.  More quantitative analysis of the effectiveness of self shielding will be summarized at the end of the background section.

\begin{figure*}
\includegraphics[height=98mm]{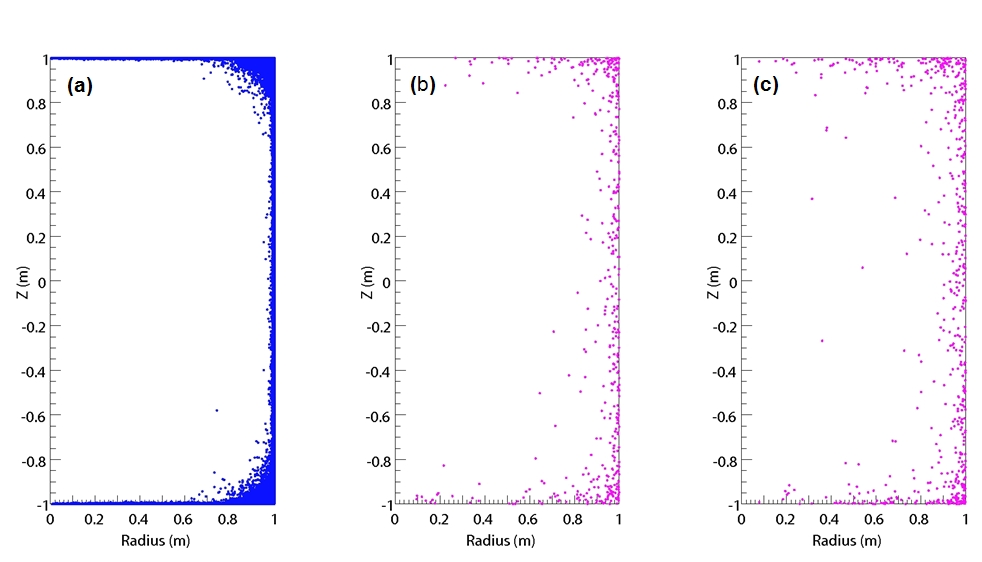}
\caption{\label{fig:epsart} Distribution of gamma ray and neutron backgrounds in 2 m diameter and 2 m high target volume. (a) 100 years worth of gamma single scatter background events after an S2/S1 cut in the case of Xe. (b) 100 years worth of neutron single scatter background events in the case of Xe. (c) The same neutron background as in (b) except for Ar.}
\end{figure*}

\subsection{Solar neutrino background}

The pp solar neutrino background (from  $\nu$-e scattering) below 20 keV electron-equivalent energy is constant at $10^{-5}$ events/keV/kg/day. This is an irreducible background.  However, because this is an electron recoil background in the S1 channel, it can be reduced by a factor of 100 from S2/S1 discrimination to the level of $10^{-7}$ events/keV/kg/day (as shown in Fig 10), or 5 events/10-ton/year.  This is a factor of two below the sensitivity corresponding to $10^{-11}$ pb WIMP-nucleon cross section, but would preclude observation of cross sections any lower than that, except by subtraction and/or by improved S2/S1 discrimination (whose electric field-dependence is still under investigation).  On the other hand, above 100 keV this background becomes observable in the S1 channel as a clear signal, which can be used by XAX to measure accurately the solar pp neutrino flux, as discussed below in \S7.

\subsection{Impurity or internal radioactivity of the noble liquid targets}

\subsubsection{Kr background in Xe}
Kr is present as an impurity in commercial Xe, with typically 50 ppb. This Kr contains $\sim10^{-11}$ cosmogenic $^{85}Kr$, which emits a continuous beta decay spectrum with a maximum energy of 690 keV and a 10-year half-life.  This would contribute a low energy beta background of $\sim$1 event/keV/kg/day to the gamma population limiting the dark matter sensitivity to $10^{-8}$ pb.  The need to remove this, to reach lower cross sections, has led to the development of reflux distillation columns which will reduce the level of Kr in Xe by a factor of 1000 per pass \cite{Rad, KRYPTON}.  Thus, using several passes, it will be possible to reduce the $^{85}Kr$ content to a negligible level.

\subsubsection{Rn background in Xe or Ar}
Radon has a lifetime of $\sim$4 days, and hence does not survive in the stored xenon gas, but can cause background problems in the Xe detectors \cite{Aln,Ang,Ben} by contaminating the target or gas systems during assembly and transferring through deposition of its decay products ($^{218}Po$, $^{214}Po$) on the detector walls where they further decay by alpha emission.  The alphas themselves are in the MeV range, and hence do not result in  low energy events if emitted into the liquid, but if emitted into the wall, the nucleus recoils into the liquid with (by momentum conservation) $< 100$ keV recoil energy and thus can produce a signal similar to that of a recoil Xe nucleus. These events can be eliminated from the data set by using the S2 position sensitivity to make a radial cut, but this cut removes a significant fraction of the target mass. The future solution must be to prevent radon from contaminating the target system, and work is in progress to understand this problem and develop procedures to eliminate it.

\subsubsection{$^{39}Ar$ background in Ar}
In the case of the Ar target, natural (atmospheric) $^{40}Ar$ contains a fraction $\sim10^{-15}$ of cosmogenically-produced $^{39}Ar$, which beta decays with a half life of 269 years and a continuous spectrum with maximum energy 570 keV, giving a total background rate $\sim$1.2/kg/s, and a low energy differential rate $\sim3\times10^{2}$ /kg/day/keV. This is too high a starting point for the two-phase discrimination technique alone, and can also produce pile-up in the data acquisition system.

The WARP collaboration \cite{Ben} has shown that pulse shape discrimination can be used to separate the nuclear recoil population from electron/gamma background above a nuclear recoil energy threshold $\sim$30 keV.  This makes it difficult to achieve a low energy threshold for liquid Ar, although there are prospects for obtaining Ar gas depleted in $^{39}Ar$ by at least a factor of 20 from an underground source \cite{Gal}, which could then be further reduced by mechanical or plasma centrifuging.  Without isotope separation, $^{39}Ar$ would produce a 1 kHz/ton accidental trigger rate. Considering that the S1-S2 time difference would be $\sim$1 ms for a multi-ton detector, such a large Ar detector could no longer tolerate the $^{39}Ar$ decay rate. However in the case of XAX, there is the additional feature of the 1\% volume position resolution of S1 allowed by the $4\pi$  QUPID array, which permits rejection of all events that do not correspond to a position coincidence of S1 and S2 within the same 1\% of the detector volume.  This effectively reduces the accidental triggers caused by an $^{39}Ar$ background by a factor of 100, allowing the XAX detector to function even in the presence of $^{39}Ar$ background higher than 1 kHz.

\subsection{Radiation from outside of the detector}
\subsubsection{Gamma rays}
Unshielded gamma rates, due to U/Th/K in the rock, are in the range $10^{5}$ - $10^{6}$ gammas/day/kg target, depending on the type of rock and level of contamination.  Assuming the higher number, a 4 m thick water shield reduces this to 0.1 gammas/kg/day in the Xe target.  Rejection by multiple scattering, combined with a 10 cm self-shielding layer (which attenuates by a factor 10 per 5 cm), further reduces this to $10^{-4}$ gammas/kg/day in the 2-20 keV electron equivalent energy range, which is then reduced by a factor of 100 discrimination via the S2/S1 cut to $10^{-6}$ gammas/kg/day,  an order of magnitude below the $10^{-5}$ gammas/kg/day background required, from Equation (3), for measurement of a $10^{-10}$ to $10^{-11}$ pb WIMP-nucleon cross section.  The above calculation makes use of the powerful self-shielding effect in the high density, high-Z, Xe target, which reduces the low energy gamma flux by one order of magnitude for each 5 cm thickness.

\subsubsection{Neutrons}
Neutrons produced from rock muons (by spallation and absorption) and U/Th (alpha-n) can be fully absorbed by the proposed 4 m water shield thickness, or by a combination of an inner water or hydrocarbon shield and an outer iron shield \cite{Bun}.  Muons closer to the detector would be rejected by a PMT Cerenkov veto around the water shield.

\begin{figure*}
\includegraphics[height=100mm]{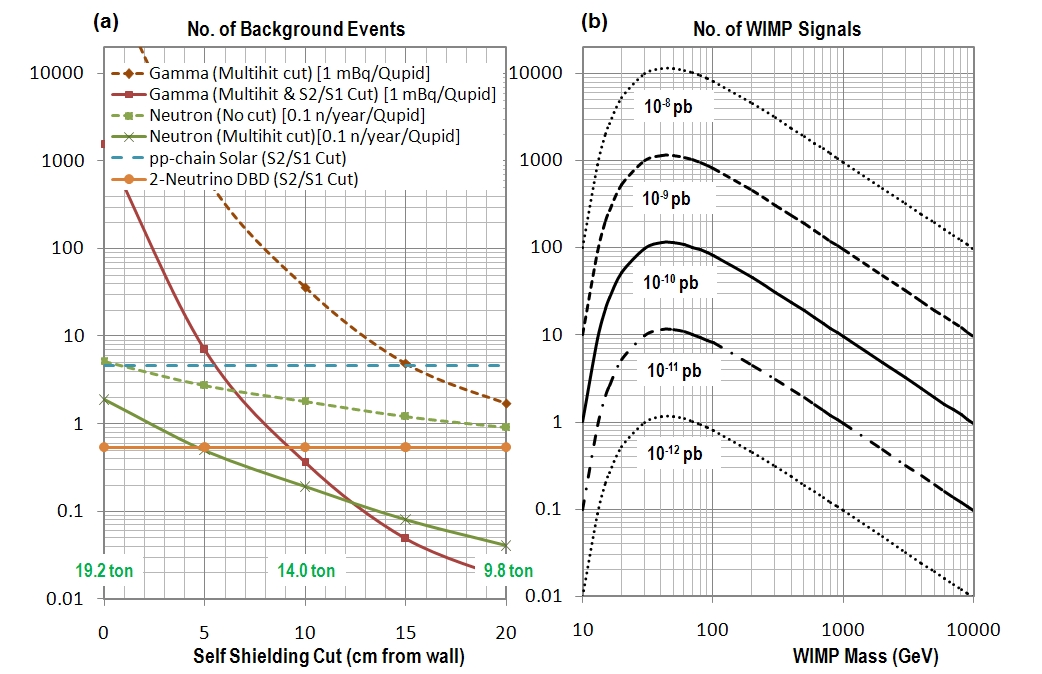}
\caption{\label{fig:epsart} (a) Expected number of backgrounds in WIMP signal range (3-15 keVee) as a function of active shielding cut (from the wall) for 10 ton-years (fiducial) of XAX data taking in the case of natural Xenon. Once a cut larger than 10 cm is performed, all the backgrounds from radioactivity disappear, except the pp-chain solar neutrino, which remains as an irreducible background at a level of 5 events/10-ton/year. (b) Expected number of WIMP signals as a function of WIMP mass.}
\end{figure*}

\subsection{Summary}

In summary, the above calculations and simulations show that the proposed XAX detector system could be constructed with sufficiently low residual background and signal identification to observe dark matter events at the rate $\sim$100 events/year for WIMP cross sections $\sim10^{-10}$ pb (for WIMP mass of $\sim$100 GeV/$c^{2}$). It would also have sufficient sensitivity for the significant detection of a signal an order of magnitude lower at $\sim10^{-11}$ pb.  

Fig 13(a) is the summary of the expected background from three major backgound sources: gamma rays from QUPIDs, neutrons from QUPIDs, and solar neutrinos. Here the expected number of backgrounds in the WIMP signal range (3-15 keVee) is plotted as a function of active shielding cut (from the wall) for 10 ton-years (fiducial) of XAX data taking in the case of natural Xenon. Once a cut larger than 10 cm is performed, all the backgrounds from radioactivity disappear, except the pp-chain solar neutrino, which remains as an irreducible background at a level of 5 events/10-ton/year.  Fig 13(b) shows the expected number of WIMP signals for 10 ton-years as a function of WIMP mass and cross section. If the cross section is $\sim10^{-11}$ pb, $\sim$10 events are expected for a WIMP mass of 30-100 GeV/$c^2$, which is above the expected 5 events from the pp solar neutrino background.  
 
Clearly this unprecedented sensitivity can be achieved only by combining (1)ultra-low radiation QUPIDs,  (2)an S2/S1 cut, (3)a multi-scattering cut and (4)a 10 cm thick self-shielding cut. It is this combination that makes the XAX detector so unique.

\section{Neutrinoless double beta decay}
The measurement of neutrino mixing, and hence non-zero mass, in large scale solar and atmospheric neutrino experiments, has provided the first indication of the neutrino mass scale, and increased the urgency of establishing whether neutrinos are Dirac or Majorana particles.  If neutrinos and antineutrinos are equivalent, then nuclei that undergo two-neutrino double beta decay must also show neutrinoless double beta decay, with the two electrons being emitted with a constant total energy.  The rate for this process is a function of nuclear matrix elements multiplied by the square of a Majorana mass parameter $m_{\nu}$, a linear combination of the neutrino mass eigenstates \cite{Avi}. For the latest measured mixing angles, $m_{\nu}$  will be in the range 0.01 - 0.1 eV for the normally ordered neutrino mass hierarchy, and 0.03 - 0.1 eV for the inverted mass hierarchy, corresponding to $0\nu\beta\beta$  decay lifetimes of $10^{27}$ - $10^{28}$ years \cite{Avi}.  This sets, for the first time, a clearly defined lifetime scale for which experiments can anticipate a non-zero signal. 

Previous experiments have set lower lifetime limits $\sim10^{24}$ - $10^{25}$ years for $0\nu\beta\beta$ decay, corresponding to upper limits of 0.3 - 1eV for the Majorana mass parameter $m_{\nu}$.   From the above estimates of the range of $m_{\nu}$, it is clear that 2-3 orders of magnitude improvement in the lifetime sensitivity is needed, with the strong likelihood of a positive observation at the $10^{27}$ - $10^{28}$ year level.   Observation of neutrinoless double beta decay would represent a major step forward in neutrino physics, and would provide:

\begin{my_enumerate1}
\item the first demonstration that neutrinos are Majorana in nature
\item the first measurement of an absolute (rather than relative) mass scale for neutrinos
\item further light on the close relationship between quark and lepton families
\end{my_enumerate1}

\begin{figure*}
\includegraphics[height=90mm]{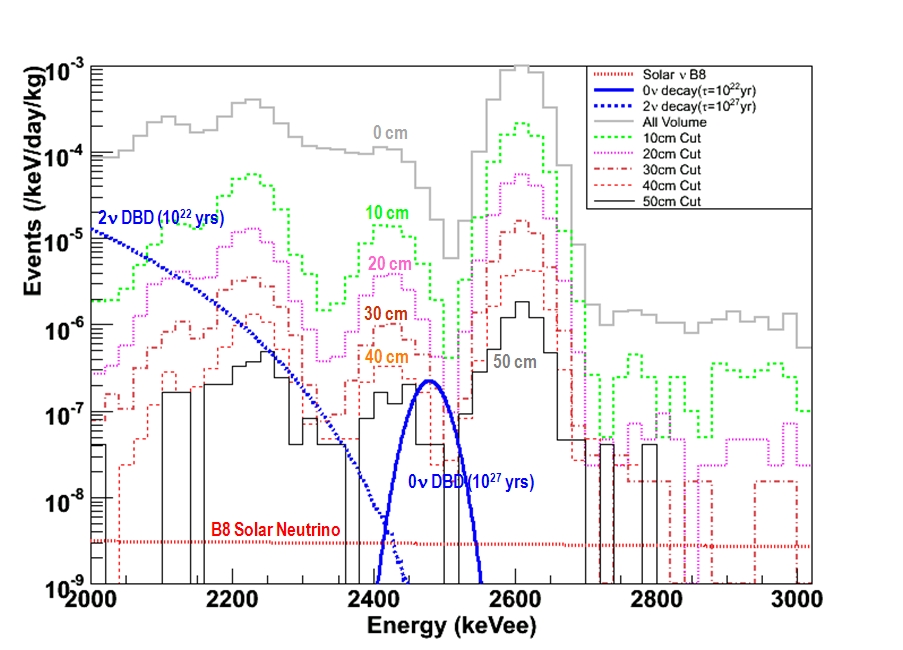}
\caption{\label{fig:epsart} Expected energy spectrum of $2\nu$/$0\nu$ double beta decays, solar neutrinos, and gamma ray backgrounds from QUPID (1 mBq per unit) as a function of self-shielding cuts. Here $^{136}Xe$ is assumed to be enriched to 80\% level. The observed energy is smeared out assuming a resolution of  $\sigma=\frac{1.5\%}{\sqrt{E\left(MeV\right)}}$.}
\end{figure*}

\begin{figure*}
\includegraphics[height=100mm]{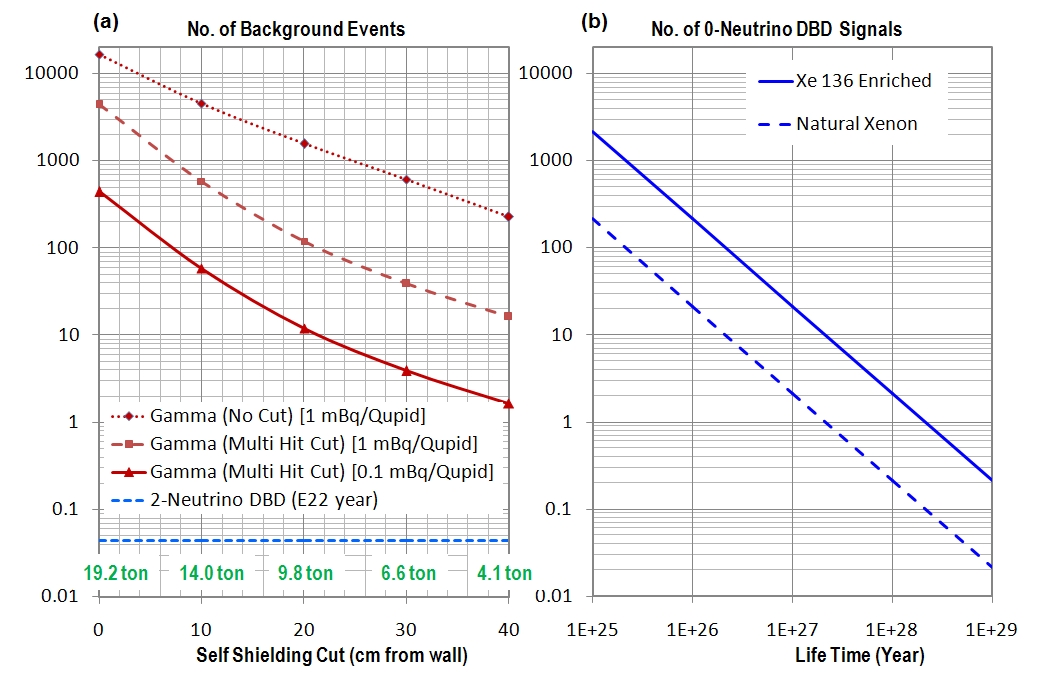}
\caption{\label{fig:epsart} (a) Expected number of backgrounds in neutrinoless double beta decay signal range ($2479 \pm 25$ keV) as a function of active shielding cut (from the wall) for 10 ton-years (fiducial) of XAX data taking.  (b) Expected number of double beta decay signals as a function of its lifetime.}
\end{figure*}

For noble gas targets, the principal double beta decay isotope is $^{136}Xe$, with the neutrinoless mode at 2479 keV. In XAX, we propose to improve the $0\nu\beta\beta$  lifetime sensitivity to $10^{27}$ - $10^{28}$ years, reaching the Majorana mass range $m_{\nu}$   $\sim$0.01 - 0.1 eV through a combination of a larger target mass and lower background than in any previous experiments. In XAX, this appears possible without needing to detect the resulting Ba nucleus, as proposed by the EXO project \cite{Exo}.  

A major advantage of combining $0\nu\beta\beta$  and dark matter searches is that the target for the latter can surround the former and provide radial self-shielding of gamma background (Fig 1 and 3) without needing to utilize a large fraction of the enriched $^{136}Xe$, as would be the case with a separate vessel.  At the same time the position sensitivity of both the S2 and S1 signals, arising from the $4\pi$  QUPID array, identifies the signals originating from the central detector, while the outer detector provides a control demonstrating the reduction of the same signals from the outer detector (or their absence, if depleted in $^{136}Xe$).

\begin{figure*}
\includegraphics[height=90mm]{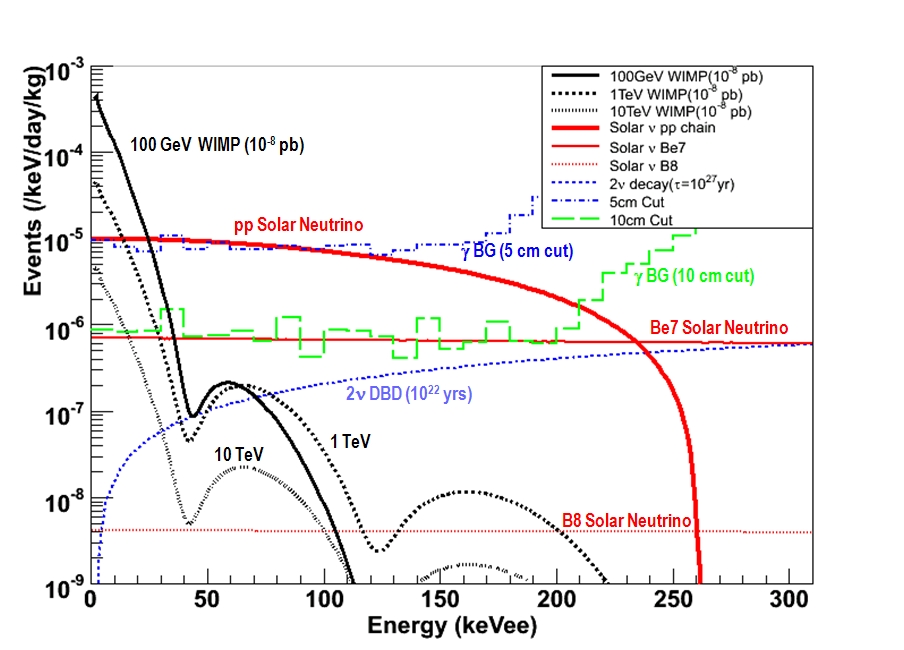}
\caption{\label{fig:epsart} Expected energy spectrum of pp solar neutrino signals, and gamma ray backgrounds from QUPID as a function of self-shielding cuts. Here $^{136}Xe$ is assumed to be depleted to 0.1\% level.}
\end{figure*}

\vspace{3mm}
Fig 14 shows the spectrum of $2\nu\beta\beta$    and $0\nu\beta\beta$    events in 80\% enriched  $^{136}Xe$, compared with the dominant photodetector gamma background, reduced in several stages (i) by a multiple scattering cut, (ii) by radially shielding with the external $^{129/131}Xe$ dark matter target, and (iii) by vertically self-shielding by sacrificing different amounts of vertical height of the target volume (using position sensitivity of the S1 signal and timing of S2 signal). Note that the number of events in the $0\nu\beta\beta$    signal region from the tail of the $2\nu\beta\beta$    spectrum is negligible. The (inner) $0\nu\beta\beta$   target is assumed to be 80\% enriched in $^{136}Xe$.  

Fig 15 summarizes the resulting background as a function of self-shielding cut (for two levels of QUPID total activity), and event rates for a $0\nu\beta\beta$  as a function of lifetime (for both natural and enriched Xe). The energy resolution at the position of the 2479 keV peak corresponds to a FWHM of 50keV (ie. $\pm1\%$) from simulations of light collection by the $4\pi$  array of photodetectors (giving $\sim4$ photoelectrons/keV).   Such a resolution can be achieved by realizing the strong anti-correlation between the S1 and S2 signals in the detector.  By combining the scintillation and ionization signals, there is a vast reduction in fluctuations of the signal compared to using scintillation or ionization alone.  An energy resolution of 1.7\% ($\sigma$) at 662 keV has been demonstrated using this technique \cite{ANTICORR}, and assuming that the resolution is proportional to $1/\sqrt{E}$ we obtain a conservative value of 1\% ($\sigma$) at 2479 keV.

Assuming the same reduction of external gamma backgrounds and Kr background to a negligible level as discussed in \S5, it can be concluded that the concentric geometry, combined with the ultra-low background of the QUPID, can ensure that a 10-event signal, corresponding to a $0\nu\beta\beta$    lifetime of $10^{27}$ years, could be reached with 5 ton-years running of the enriched target. If an initial trial period is performed by natural Xe in the central target region,  the sensitivity would be a factor of 10 lower (since natural Xe contains 8.9\% $^{136}Xe$), but nevertheless capable of a preliminary detection at the $10^{26}$-$10^{27}$ year level, which could subsequently be confirmed and enhanced by substitution of a fully enriched target.

\section{Detection of pp solar neutrinos}
Up to the present time, the largest (99\%) component of the solar neutrino flux, the pp-chain, has not yet been measured. The pp neutrino spectrum extends from 260 keV down to zero, and has already been considered above as a possible low energy background to dark matter experiments, but suppressed to a low level by a factor of 100 S2/S1 discrimination of the two phase liquid Xe.  In case of detection of the pp solar neutrinos, however, this suppression would not be required and its spectrum emerges from the dark matter spectrum and becomes a measurable (neutrino-electron scattering) signal in the Xe target \cite{SOLARNU}.   

This would be true also for the Ar target (which has the same pp solar neutrino rate per ton) were it not for the $^{39}Ar$ beta background at a level of $3\times10^{2}$/kg/d/keV, extending to 570 keV, compared with the pp solar neutrino rate $10^{-5}$/kg/d/keV, extending to 260 keV. Thus even with significant depletion of the $^{39}Ar$ it would still swamp the whole of the pp neutrino spectrum.  However, either the very efficient removal of $^{39}Ar$ or the previously-mentioned option of using Ne as a dark matter target in the second vessel would restore the possibility of using this as a second pp solar neutrino target, with the same expected rate per kg from electron scattering \cite{CLEAN1}.

\begin{figure}
\includegraphics[height=95mm]{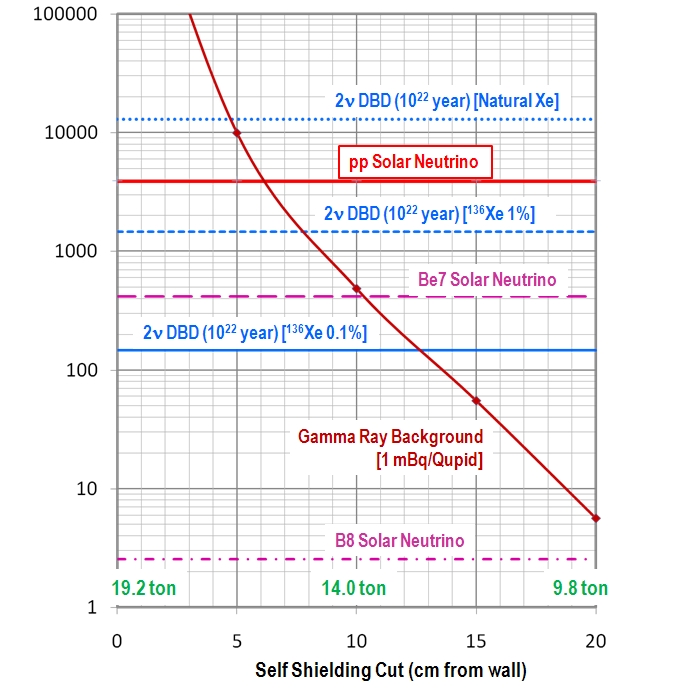}
\caption{\label{fig:epsart} Expected number of backgrounds in pp solar neutrino signal range (30-200keV) as a function of active shielding cut (from the wall) for 10 ton-years (fiducial) of XAX data taking.}
\end{figure}

The normal two-neutrino double beta decay rates from the 8.9\% $^{136}Xe$ in natural Xe exceed those of the pp solar neutrinos above 30 keV. Thus for the outer Xe to be used for solar neutrino detection, it is necessary that the $^{136}Xe$ component be depleted by a factor of 100 \cite{solar}, \cite{SOLARNU}. This may not represent an additional isotope separation requirement, because the enhancement of $^{136}Xe$ for the inner ($0\nu\beta\beta$)  target would produce the depleted natural Xe as an automatic by-product.

\vspace{3mm}
Fig 16 compares the absolute rates and energy spectra of pp solar neutrinos, two-neutrino double beta decay (from Xenon with $^{136}Xe$ depleted to 0.1\%), and hypothetical dark matter fluxes (for 100 GeV, 1 TeV, and 10 TeV mass particles and $10^{-8}$ pb cross section).  As in the case of the gamma background to the $0\nu\beta\beta$  decay signal, an excellent position resolution from the double-phase TPC operation allows us to use the self-shielding by the outer detector layers to reduce the gamma background to an order of magnitude or more below the pp neutrino rates.

Fig 17 summarizes the results of simulations giving the number of gamma background events/10-ton/year, in a 30-200 keV window, as a function of distance from the outer surface of the target, compared with the pp and $^{7}Be $ solar neutrino rates \cite{solar} in natural and $^{136}Xe$-depleted Xe.   Taking the example of a 10 cm self-shielding cut and Xe depleted to 0.1\% $^{136}Xe$, 4,000 pp events could be collected with a 10 ton-year exposure, enabling the pp flux to be measured to an accuracy of $\pm2\%$. With Ne substituted in the Ar vessel, and a 10 ton-year exposure in both, the precision could be improved to $\pm1\%$.

\section{Conclusions}
We have studied the capabilities of XAX, a multi-ton, three-target detection system capable of achieving new levels of sensitivity in the detection and measurement of dark matter, neutrinoless double beta decay and pp solar neutrino signals. The three targets are independent but symbiotic. Two liquid Xe targets, one enriched and the other depleted in $^{136}Xe$, are arranged concentrically, allowing the outer $^{129/131}Xe$ to act both as a detector for dark matter and solar neutrinos, while providing additional shielding for the detection of neutrinoless double beta decay in the inner $^{136}Xe$.  A separate vessel of similar or larger total size houses a liquid Ar (or optionally liquid Ne) target which would provide an additional and confirmatory pp solar neutrino detector, but crucially will identify a genuine dark matter WIMP signal through the coherent $A^{2}$ dependence of a low energy cross section, which thus gives an order of magnitude larger signal in Xe than in Ar.  

A key feature of the proposed system is the light collection and position sensitivity achievable by means of a $4\pi$  array of photodetectors, together with the extremely low background levels made possible by the QUPID - a new all-quartz hybrid avalanche photodiode now under development jointly by UCLA and Hamamatsu Photonics.

The system includes both zero spin and non-zero spin Xe dark matter targets, and by inclusion of Ar (optionally Ne), provides two values of atomic number and nuclear mass for signal identification and, with sufficient events, measurement of the WIMP mass.  We have shown that background levels can be reduced or rejected to a level sufficient to identify a dark matter signal at a WIMP-nucleon cross section level as low as $10^{-11}$ pb, and to measure a dark matter energy spectrum of $> 100$ events/year at cross section $>10^{-10}$ pb, thus covering a large fraction of the favored parameter space of supersymmetry theory.

\vspace{3mm}
For the $^{136}Xe$, target, we have shown that background levels can be reduced sufficiently to reach a $0\nu\beta\beta$    half life sensitivity of $\sim10^{27}$ years if an individual QUPID activity of 1 mBq  is achieved, and  $\sim10^{28}$ years for an individual QUPID activity of 0.1 mBq.  Thus for the first time the vital goal of positive detection of a Majorana mass level of 0.1 - 0.01 eV, in agreement with the mass level suggested by neutrino mixing results, would be achievable in XAX.

With the background reductions necessary for dark matter and $0\nu\beta\beta$    decay, the detection of the pp solar neutrino flux becomes a by-product of the XAX detection system, provided that the outer Xe target is depleted in $^{136}Xe$ to reduce the $2\nu\beta\beta$    background.  However, again the gamma background is pivotal, and the position sensitivity and ultra-low background of the QUPID array here plays an essential part, since it allows a software-controlled amount of self shielding to be applied to the outer $^{129/131}Xe$ detector, while the latter independently retains its function as a full shield for the inner $^{136}Xe$.  A similar software position-sensitive control of the self-shielding in the Ar or optional Ne target will play a part in optimizing dark matter and pp solar neutrino detection in that target also.  With the foreseeable background reduction, we have shown that the dominant pp solar neutrino flux can be measured to an accuracy of 1-2\%.

\vspace{3mm}
We have shown that three of the most important particle physics and astrophysics goals can be achieved in XAX, with the component detectors at the 10 ton level.  Clearly, it is also the case that considerable advances in sensitivity can be achieved by a similar configuration with component detectors at the one ton level, although the difference is more than just the factor of 10 in mass, since the self shielding depends on absolute thickness and is less effective at lower masses.  Thus it is at the 10 ton level that the most dramatic advances in detection capability become possible, as demonstrated by the results shown throughout this paper.

\vspace{5mm}
\noindent
{\bf Acknowledgements}

\vspace{2mm}

The concept of achieving three goals through the use of Liquid Xe was originally inspired by Y. Suzuki and others in the XMASS collaboration.  We are especially indebted to M. Nakahata in XMASS for providing us with his code for computing the solar neutrino flux and interactions with Xe.  
QUPID is the result of extensive discussion and collaboration with Hamamatsu Photonics Co., in particular M. Suyama.  We also thank J. Takeuchi, T. Hakamata, S. Muramatsu and A. Fukasawa at Hamamatsu for their continuous support and development.
We are thankful for useful discussions with E. Aprile, K. Giboni, K. Ni, R. Cousins, B. Sadoulet, H. Sobel, R. Gaitskell, T. Shutt, D. McKinsey, F. Calaprice and C. Galbiati.
This work was supported in part by US DOE grant DE-FG-03-91ER40662 , and by NSF grants PHY-0139065/PHY-0653459 as well as the NSF REU program.
This work was also supported in part by discretionary funds from UCLA Vice Chancellor R. Peccei, Dean J. Rudnick and Department Chair F. Coroniti. 
Lastly, the authors would especially like to thank PK Williams for his strong support and encouragement over the last two decades during his tenure at DOE.

\bibliography{XAX}

\begin{thebibliography}{26}
\expandafter\ifx\csname natexlab\endcsname\relax\def\natexlab#1{#1}\fi
\expandafter\ifx\csname bibnamefont\endcsname\relax
  \def\bibnamefont#1{#1}\fi
\expandafter\ifx\csname bibfnamefont\endcsname\relax
  \def\bibfnamefont#1{#1}\fi
\expandafter\ifx\csname citenamefont\endcsname\relax
  \def\citenamefont#1{#1}\fi
\expandafter\ifx\csname url\endcsname\relax
  \def\url#1{\texttt{#1}}\fi
\expandafter\ifx\csname urlprefix\endcsname\relax\def\urlprefix{URL }\fi
\providecommand{\bibinfo}[2]{#2}
\providecommand{\eprint}[2][]{\url{#2}}

\bibitem[{\citenamefont{Alner et~al.}(2007)}]{Aln}
\bibinfo{author}{\bibfnamefont{G.~J.} \bibnamefont{Alner}}
  \bibnamefont{et~al.}, \bibinfo{journal}{Astroparticle Physics}
  \textbf{\bibinfo{volume}{28}}, \bibinfo{pages}{287} (\bibinfo{year}{2007}).

\bibitem[{\citenamefont{Angle et~al.}(2008)}]{Ang}
\bibinfo{author}{\bibfnamefont{J.}~\bibnamefont{Angle}} \bibnamefont{et~al.},
  \bibinfo{journal}{Phys. Rev. Lett.} \textbf{\bibinfo{volume}{100}}
  (\bibinfo{year}{2008}).

\bibitem[{\citenamefont{Benetti et~al.}(2008)}]{Ben}
\bibinfo{author}{\bibfnamefont{P.}~\bibnamefont{Benetti}} \bibnamefont{et~al.},
  \bibinfo{journal}{Astroparticle Physics} \textbf{\bibinfo{volume}{28}},
  \bibinfo{pages}{495} (\bibinfo{year}{2008}).

\bibitem[{\citenamefont{Angle et~al.}(2007)}]{XeTPC}
\bibinfo{author}{\bibfnamefont{J.}~\bibnamefont{Angle}} \bibnamefont{et~al.},
  \bibinfo{journal}{Nucl. Phys. B-Proc. Suppl.} \textbf{\bibinfo{volume}{173}},
  \bibinfo{pages}{117} (\bibinfo{year}{2007}).

\bibitem[{\citenamefont{Ereditato et~al.}(2006)}]{ArTPC}
\bibinfo{author}{\bibfnamefont{A.}~\bibnamefont{Ereditato}}
  \bibnamefont{et~al.}, \bibinfo{journal}{Nucl. Phys. B-Proc. Suppl.}
  \textbf{\bibinfo{volume}{154}}, \bibinfo{pages}{163} (\bibinfo{year}{2006}).

\bibitem[{\citenamefont{Brearley et~al.}(1982)\citenamefont{Brearley, Bore,
  Evans, and Scott}}]{PSD}
\bibinfo{author}{\bibfnamefont{I.~R.} \bibnamefont{Brearley}},
  \bibinfo{author}{\bibfnamefont{A.}~\bibnamefont{Bore}},
  \bibinfo{author}{\bibfnamefont{N.}~\bibnamefont{Evans}}, \bibnamefont{and}
  \bibinfo{author}{\bibfnamefont{M.~C.} \bibnamefont{Scott}},
  \bibinfo{journal}{Nucl. Inst. Meth.} \textbf{\bibinfo{volume}{192}},
  \bibinfo{pages}{439} (\bibinfo{year}{1982}).

\bibitem[{\citenamefont{SNO-Collaboration}(2000)}]{SNO}
\bibinfo{author}{\bibnamefont{SNO-Collaboration}}, \bibinfo{journal}{Nucl.
  Inst. Meth.}  (\bibinfo{year}{2000}).

\bibitem[{\citenamefont{Brunetti et~al.}(2005)}]{WARP1}
\bibinfo{author}{\bibfnamefont{R.}~\bibnamefont{Brunetti}}
  \bibnamefont{et~al.}, \bibinfo{journal}{New Astr. Rev.}
  \textbf{\bibinfo{volume}{40}}, \bibinfo{pages}{265} (\bibinfo{year}{2005}).

\bibitem[{\citenamefont{McKinsey and Coakley}(2005)}]{CLEAN1}
\bibinfo{author}{\bibfnamefont{D.~N.} \bibnamefont{McKinsey}} \bibnamefont{and}
  \bibinfo{author}{\bibfnamefont{K.~J.} \bibnamefont{Coakley}},
  \bibinfo{journal}{Astroparticle Physics} \textbf{\bibinfo{volume}{22}},
  \bibinfo{pages}{355} (\bibinfo{year}{2005}).

\bibitem[{\citenamefont{Abe et~al.}(2008{\natexlab{a}})}]{XMASS}
\bibinfo{author}{\bibfnamefont{K.}~\bibnamefont{Abe}} \bibnamefont{et~al.},
  \bibinfo{journal}{J. Phys.: Conf. Ser.} \textbf{\bibinfo{volume}{120}}
  (\bibinfo{year}{2008}{\natexlab{a}}).

\bibitem[{\citenamefont{Coakley and McKinsey}(2004)}]{CLEAN}
\bibinfo{author}{\bibfnamefont{K.~J.} \bibnamefont{Coakley}} \bibnamefont{and}
  \bibinfo{author}{\bibfnamefont{D.~N.} \bibnamefont{McKinsey}},
  \bibinfo{journal}{Nucl. Inst. Meth.} \textbf{\bibinfo{volume}{A 522}},
  \bibinfo{pages}{504} (\bibinfo{year}{2004}).

\bibitem[{\citenamefont{Ferrari et~al.}(2006)}]{WARP}
\bibinfo{author}{\bibfnamefont{N.}~\bibnamefont{Ferrari}} \bibnamefont{et~al.},
  \bibinfo{journal}{J. Phys.: Conf. Ser.} \textbf{\bibinfo{volume}{39}},
  \bibinfo{pages}{111} (\bibinfo{year}{2006}).

\bibitem[{\citenamefont{Arisaka}(2000)}]{Ari1}
\bibinfo{author}{\bibfnamefont{K.}~\bibnamefont{Arisaka}},
  \bibinfo{journal}{Nucl. Inst. Meth.} \textbf{\bibinfo{volume}{A442}},
  \bibinfo{pages}{80} (\bibinfo{year}{2000}).

\bibitem[{\citenamefont{Arisaka and Wang}()}]{Ari2}
\bibinfo{author}{\bibfnamefont{K.}~\bibnamefont{Arisaka}} \bibnamefont{and}
  \bibinfo{author}{\bibfnamefont{H.}~\bibnamefont{Wang}}, \bibinfo{note}{to be
  published to NIM, US patent pending}.

\bibitem[{\citenamefont{Lewin and Smith}(1996)}]{Lew}
\bibinfo{author}{\bibfnamefont{J.~D.} \bibnamefont{Lewin}} \bibnamefont{and}
  \bibinfo{author}{\bibfnamefont{P.~F.} \bibnamefont{Smith}},
  \bibinfo{journal}{Astroparticle Physics} \textbf{\bibinfo{volume}{6}},
  \bibinfo{pages}{87} (\bibinfo{year}{1996}).

\bibitem[{\citenamefont{Roszkowski et~al.}(2007)}]{SUSY}
\bibinfo{author}{\bibfnamefont{L.}~\bibnamefont{Roszkowski}}
  \bibnamefont{et~al.}, \bibinfo{journal}{Journal of High Energy Physics 07}
  \textbf{\bibinfo{volume}{075}} (\bibinfo{year}{2007}).

\bibitem[{\citenamefont{Smith and Lewin}(1990)}]{Smith}
\bibinfo{author}{\bibfnamefont{P.~F.} \bibnamefont{Smith}} \bibnamefont{and}
  \bibinfo{author}{\bibfnamefont{J.~D.} \bibnamefont{Lewin}},
  \bibinfo{journal}{Physics Reports} \textbf{\bibinfo{volume}{187}},
  \bibinfo{pages}{203} (\bibinfo{year}{1990}).

\bibitem[{\citenamefont{Bungau et~al.}(2005)}]{Bun}
\bibinfo{author}{\bibfnamefont{C.}~\bibnamefont{Bungau}} \bibnamefont{et~al.},
  \bibinfo{journal}{Astroparticle Physics} \textbf{\bibinfo{volume}{23}},
  \bibinfo{pages}{97} (\bibinfo{year}{2005}).

\bibitem[{\citenamefont{Takeuchi}(2005)}]{Rad}
\bibinfo{author}{\bibfnamefont{Y.}~\bibnamefont{Takeuchi}},
  \bibinfo{journal}{AIP Conf. Proc.} \textbf{\bibinfo{volume}{785}},
  \bibinfo{pages}{130} (\bibinfo{year}{2005}).

\bibitem[{\citenamefont{Abe et~al.}(2008{\natexlab{b}})}]{KRYPTON}
\bibinfo{author}{\bibfnamefont{K.}~\bibnamefont{Abe}} \bibnamefont{et~al.},
  \bibinfo{journal}{{``}Distillation of Liquid Xenon to Remove Krypton{"}}
  (\bibinfo{year}{2008}{\natexlab{b}}), \eprint{arXiv:0809.4413v2}.

\bibitem[{\citenamefont{Galbiati et~al.}(2008)}]{Gal}
\bibinfo{author}{\bibfnamefont{C.}~\bibnamefont{Galbiati}}
  \bibnamefont{et~al.}, \bibinfo{journal}{J. Phys.: Conf. Ser.}
  \textbf{\bibinfo{volume}{120}} (\bibinfo{year}{2008}).

\bibitem[{\citenamefont{Avignone et~al.}(2005)}]{Avi}
\bibinfo{author}{\bibfnamefont{F.~T.} \bibnamefont{Avignone}}
  \bibnamefont{et~al.}, \bibinfo{journal}{New Journal of Physics}
  \textbf{\bibinfo{volume}{7}}, \bibinfo{pages}{6} (\bibinfo{year}{2005}).

\bibitem[{\citenamefont{Danilov et~al.}(2000)}]{Exo}
\bibinfo{author}{\bibfnamefont{M.}~\bibnamefont{Danilov}} \bibnamefont{et~al.},
  \bibinfo{journal}{Phys. Lett. B} \textbf{\bibinfo{volume}{480}},
  \bibinfo{pages}{12} (\bibinfo{year}{2000}).

\bibitem[{\citenamefont{Aprile et~al.}(2007)}]{ANTICORR}
\bibinfo{author}{\bibfnamefont{E.}~\bibnamefont{Aprile}} \bibnamefont{et~al.},
  \bibinfo{journal}{Phys. Rev. B} \textbf{\bibinfo{volume}{76}},
  \bibinfo{pages}{014115} (\bibinfo{year}{2007}).

\bibitem[{\citenamefont{Suzuki}(2000)}]{SOLARNU}
\bibinfo{author}{\bibfnamefont{Y.}~\bibnamefont{Suzuki}},
  \bibinfo{journal}{{``}Low Energy Solar Neutrino Detection by Using Liquid
  Xenon{"}}  (\bibinfo{year}{2000}), \eprint{arXiv:hep-ph/0008296}.

\bibitem[{\citenamefont{Suzuki}(2005)}]{solar}
\bibinfo{author}{\bibfnamefont{Y.}~\bibnamefont{Suzuki}},
  \bibinfo{journal}{Nucl. Phys. B-Proc. Suppl.} \textbf{\bibinfo{volume}{143}},
  \bibinfo{pages}{27} (\bibinfo{year}{2005}).

\end{thebibliography}
\end{document}